\documentclass{emulateapj}
\usepackage{natbib,graphicx,amsmath,amsthm,ulem,color}

\def\pomega{\varpi}

\newcommand{\be}{\begin{equation}}
\newcommand{\ee}{\end{equation}}
\newcommand{\beqn}{\begin{eqnarray}}
\newcommand{\eeqn}{\end{eqnarray}}
\newcommand{\bi}{\begin{itemize}}
\newcommand{\ei}{\end{itemize}}

\def\refnew#1{(\ref{#1})}

\def\pomega{\varpi}

\def\icarus{{Icarus}}

\usepackage{ulem}

\begin{document} 

\title{Resonant Repulsion of Kepler Planet Pairs}	

\author{Yoram Lithwick\altaffilmark{1} and Yanqin Wu\altaffilmark{2}}
\altaffiltext{1}{Dept. of Physics and Astronomy, Northwestern University, 2145 Sheridan Rd., Evanston, IL 60208
\& Center for Interdisciplinary Exploration and Research in Astrophysics (CIERA)}
 \altaffiltext{2}{Department of Astronomy and Astrophysics, University of Toronto, Toronto, ON M5S 3H4, Canada}
	
\begin{abstract}
  Planetary systems discovered by the Kepler space telescope exhibit
  an intriguing feature. While the period ratios of adjacent low-mass
  planets appear largely random, there is a significant excess of
  pairs that lie just wide of resonances and a deficit on the near
  side. We demonstrate that this feature naturally arises when two
  near-resonant planets interact in the presence of weak dissipation
  that damps eccentricities.  The two planets repel each other as
  orbital energy is lost to heat.  This moves near-resonant pairs just
  beyond resonance, by a distance that reflects the integrated
  dissipation they experienced over their lifetimes. We find that the
  observed distances may be explained by tides  if 
        tidal dissipation is unexpectedly efficient (tidal
  quality factor $\sim 10$).  Once the effect of resonant repulsion is
  accounted for, the initial orbits of these low mass planets show
  little preference for resonances.  This could constrain
   their origin.
 \end{abstract}

\section{Introduction}

NASA's Kepler mission is revolutionizing our knowledge of planetary
systems.  It has already discovered thousands of transiting planetary
candidates, including hundreds of systems with two or more planets
\citep{Batalhaetal12}. Most of these are Neptune- or Earth-sized
planets.  To date, one of the most intriguing Kepler discoveries is
that, while the spacing between  planets appears to be roughly
random, there is a distinct excess of planetary pairs just wide of
certain resonances, and a nearly empty gap just narrow of them
\citep{Lissaueretal11,Fabryckyetal12}.  These features are
particularly prominent near the 3:2 and 2:1 resonances, and
affects
planets that fall within a few percent of resonances
\citep{Fabryckyetal12}.

Is this resonance asymmetry a feature planetary systems are born with,
or one  they acquire much later on?  Many studies have reported
that planets become trapped into first-order resonances when they
migrate in  protoplanetary disks
\citep[e.g.,][]{LeePeal2003,Snellgrove2001,Papaloizou2005}. In fact,
the presence of resonances among giant planets detected by radial
velocity has been regarded as  strong evidence for  disk
migration  \citep[e.g.,][]{marcy2001,tinney}. However,
Kepler's low-mass planets appear to be less influenced by resonances,
and the pile-ups just outside resonances are partly counterbalanced by
the gaps  inside them.

In this paper, we identify a process that can modify the pair
separation and give rise to the observed resonance asymmetry. 
But first, let us consider
 a commonly invoked mechanism, tidal circularization.  If
the inner planet is eccentric, tides raised on it would damp its
eccentricity, decrease its semi-major axis, and hence increase the
period ratio of the pair \citep{Novak, Terquem}\footnote{Tides raised
  on the star play little role.}.  Adopting the equilibrium tide
expression from \cite{Hut}, the damping rate for a psudo-synchronized
planet is
\begin{eqnarray}
\gamma_e = {1\over e}{{de}\over{dt}}  = 
-{{9}\over 2}\, {{ k_{2}}\over{ T_1}} q(1+q) \left({{R_1}\over
a}\right)^8 ,
\label{eq:dedtide}
\end{eqnarray}
where $q = M_*/m_1$ is the mass ratio of the star to planet, $k_2$ the
tidal love number, $R_1$ the inner planet's radius and $a$ its orbital
separation. In this tidal model, $T_1 = {{R_1^3}/({G m_1 \tau_1})}$
where $\tau_1$ is the assumed constant tidal lag time which we take to
be $\tau_1 = P_1 / (2 Q_1 )$, with $Q_1$ the inner planet's tidal quality factor
\citep{GoldreichSoter} and $P_1$ its orbital period.  Numerically,
\begin{eqnarray}
\gamma_e & \sim &
(6.5\times 10^7 \, {\rm yrs})^{-1}\,
\left({Q_1}\over{10}\right)^{-1} 
\left({k_{2}}\over{0.1}\right) \,
\left({{M_*}\over{M_\odot}}\right)^{-2/3}\, \nonumber \\ & & \times
 \left({{m_1}\over{10 M_\oplus}}\right)^{-1}\,
\left({{R_1}\over{2R_\oplus}}\right)^{5}\,
\left({{P_1}\over{5 {\rm day}}}\right)^{-13/3}.
\label{eq:taue}
\end{eqnarray}
 The orbital
decay rate is   ${\dot a}/a = 2 e {\dot e}$
because orbital angular momentum
is largely conserved (assuming
  that $e\ll 1$).
So tidal evolution could have potentially circularized orbits inward
of $\sim 10$ days. As it does so, it moves the inner planet inward by
$\delta a/a \sim - e_1^2$. This increases the period ratio for a
planet pair by a fractional amount of $3e_1^2/2 = 1.5\%
(e_1/0.1)^2$. 
However, tidal circularization alone can not reproduce the observed
asymmetry: assuming all near-resonant pairs were initially uniformly
distributed in their period ratios, all systems march to larger period
ratios by a comparable amount. This produces neither gap nor peak.

A more selective mechanism is required.  In this paper, we show
that for a pair of planets that happen to lie near a mean-motion
resonance, dissipation causes the planets to repel each other.  The
rate of repulsion is  greatest at exact resonance and falls off
steeply away from resonance.  Planets that are initially slightly
closer than resonance are pushed wide of the resonance; those that are
initially wider are pushed even further apart. And planet pairs far
away from the resonance are not affected. So the combined action of
resonant interaction and damping naturally give rise to the observed
resonance asymmetry.  This effect, which we term ``resonant
repulsion,''
 was  investigated by 
\cite{greenberg} for the Galilean satellites; by
\citet{lwpluto} to possibly account for
the orbits of Pluto's minor moons; 
 and by
\citet{johnpap2011} 
for multiple planet systems.
\cite{bm} independently arrived at many of the results presented
in this paper;  their paper was posted to arxiv.org at the same time
as this one.

\section{Resonant Repulsion}
\label{sec:analytic}

We consider the evolution of two planets orbiting a star and assume that
 the interaction 
between the planets is predominantly due to the 2:1 resonance. We will also 
include  weak external eccentricity-damping forces. 
The energy (or Hamiltonian) of the two planets is, to leading order in eccentricity,
\begin{eqnarray}
H &=& -{GM_* m_1\over 2a_1}-{GM_* m_2\over 2a_2} 
-{Gm_1m_2\over a_2}\times  \nonumber
\\ 
&& 
\!\!\!\!\!\!\!\!\!\!\!\!\!\!\!\!\!
 \left(f_1e_1\cos\left(2\lambda_2-\lambda_1-\pomega_1\right)
+f_2e_2\cos\left(2\lambda_2-\lambda_1-\pomega_2\right)\right)  \ \ 
\label{eq:hamil1}
\end{eqnarray}
where we follow standard notation \citep[e.g.,][]{MD00}, with the
orbital parameters for the inner planet denoted by
$\{a_1,e_1,\lambda_1,\pomega_1\}$, and those for the outer planet
subscripted by 2.  The mass of the star and planets are $M_*,m_1,m_2$,
and the Laplace coefficients are $f_1=-(2+\alpha D/2)b_{1/2}^2$ and
$f_2=(3/2+\alpha D/2)b_{1/2}^1-2\alpha$ \citep{MD00}.  Near 2:1
resonance ($\alpha=2^{-2/3}$), the Laplace coefficients are $f_1 =
-1.19$ and $f_2 = 0.428$.

We  choose units such that
\be
GM_*=1 \ ,
\ee
and assume that the eccentricities are small.
In terms of the complex eccentricity
\be
z_j \equiv e_je^{i\pomega_j} \ ,
\ee
the equations of motion for planet $j$ are
\citep[e.g.][]{MD00,lwpluto}
\beqn
{d\lambda_j\over dt}&=&{2\sqrt{a_j}\over m_j}{\partial H\over\partial a_j} \\
{dz_j\over dt}&=&-{2i\over m_j\sqrt{a_j}}{\partial H\over \partial z_j^*} \\
{da_j\over dt} &=& -{2\sqrt{a_j}\over m_j}{\partial H\over\partial \lambda_j} 
\eeqn

To leading order in $m_j/M_*$, the semi-major axes are constant, and 
the equations for $\lambda_j$ are  
\be
{d\lambda_j\over dt}=n_j \ ,
\ee
where
\be
n_j\equiv a_j^{-3/2} \ .
\ee
Hence
\be
\lambda_j\approx n_j t \ ,
\ee

The eccentricity equations become, after adding damping terms,
\beqn
{dz_1\over dt} &=& {i\mu_2n_2 }\sqrt{a_2\over a_1}f_1e^{i\phi}-\gamma_{e1} z_1 
\label{eq:z1}
\\
{dz_2\over dt}&=&{i\mu_1}n_2f_2 e^{i\phi}-\gamma_{e2} z_2 \ ,
\label{eq:z2}
\eeqn
where
\beqn
\mu_j&\equiv& m_j/M_* \\
\phi &\equiv& 2\lambda_2-\lambda_1 \approx -2\Delta\cdot n_2t \ .
\eeqn
Here
\be
\Delta\equiv {n_1-2n_2\over 2n_2}  \ .
\label{eq:deltadef}
\ee is the fractional distance to nominal resonance.  When $\Delta<0$
the pair is on the near side of resonance, otherwise it is on the far
side.\footnote{ For brevity, we often refer to nominal resonance
  ($\Delta=0$) as simply resonance. This should not be
  confused with a pair being locked in resonance, i.e. in a state
  where the resonant angles librate.} The $\gamma_{ej}$ in Equations
(\ref{eq:z1})--(\ref{eq:z2}) denote the eccentricity damping rates on
each of the two planets due to some external force (e.g., tides or a
dissipative disk).  We assume that $\gamma_{ej}\ll |\Delta n_2|$.

We discard the free solutions
 to Equations (\ref{eq:z1})--(\ref{eq:z2})
 because they decay to zero at the  rates
$\gamma_{ej}$,  much faster than the rate of semi-major axis evolution, as we shall see below.
The forced eccentricities are, to first order
in $\gamma_{ej}/(\Delta n_2)\ll 1$:
\beqn
z_1&=&-{\mu_2\over 2\Delta} f_1 \sqrt{a_2\over a_1}e^{i\phi}\left( 
 1-i{\gamma_{e1}\over 2\Delta n_2}\right) \label{eq:efo1} \\ 
 \nonumber \\
 z_2&=&-{\mu_1\over 2\Delta}f_2 e^{i\phi}\left(
 1-i{\gamma_{e2}\over 2\Delta n_2}\right) \ . \label{eq:efo2}
\eeqn
The
 small phase shift, $O(\gamma_{ej})$, relative to the undamped forced
eccentricities  plays a crucial role in resonant repulsion.

Inserting the above forced eccentricities into the semi-major axis
equations yields, as in \cite{lwpluto}, \beqn {d\ln a_1\over dt}&=&-
{\beta\over 2}{\mu_1^2\over\Delta^2}(\gamma_{e1}
f_1^2\beta+\gamma_{e2} f_2^2)
-\gamma_{a1}|z_1|^2. \label{eq:dlna}  \\
{d\ln a_2\over dt}&=& {\mu_1^2\over\Delta^2}(\gamma_{e1}
f_1^2\beta+\gamma_{e2} f_2^2) \ -\gamma_{a2}|z_2|^2 \label{eq:dlna2}
\eeqn where \beqn \beta&\equiv& {\mu_2\sqrt{a_2}\over\mu_1\sqrt{a_1}}
\eeqn and we have included additional damping terms with rates
$\gamma_{aj}e_j^2$.  This form for the damping rate is applicable for
any process that conserves angular momentum, such as
tides (see below).  By contrast, if the planet is migrated in a disk,
or pushed by tides raised on the central body (as for Jupiter's
moons), the induced rate of change of $a$ would be independent of
eccentricity.  We shall not consider those kinds of forces.

We conclude that the distance to resonance changes at the rate
\begin{eqnarray}
 {{d\Delta}\over{dt}} =    {3\over 4}
  {{\mu_1}^2\over{\Delta^2}} \Gamma  \ ,
\label{eq:ddelta}
  \end{eqnarray}
 where
  \begin{eqnarray}
  \Gamma\equiv (2 + \beta)(\gamma_{e1}
    f_1^2 \beta + \gamma_{e2} f_2^2)
   + {\gamma_{a1} f_1^2 \beta^2 -
    \gamma_{a2} f_2^2\over 2} \ ,
    \label{eq:Gam}
 \end{eqnarray}
 for $|\Delta|\ll 1$.
We verify this rate with an N-body simulation below.

As long as $\Gamma>0$, as we shall argue is the case, 
then $\Delta$ always increases, independent of the sign of $\Delta$.  
A pair of planets that is initially spaced closer than nominal resonance ($\Delta<0$)
 will tend to be pushed outside of resonance, i.e. to $\Delta>0$. 
 And a pair initially outside of resonance will be pushed even further apart.
 We term this effect
{\it resonant repulsion}. 
 Furthermore, since the speed of migration is  slowest far from resonance, 
 the region near nominal resonance ($\Delta = 0$) should
be unoccupied, and resonant pairs should evacuate  the resonance region
and pile up outside.
This
 will  lead to an asymmetry, with more planets outside of nominal resonance than inside.

\begin{figure}
\centerline{\includegraphics[width=0.49\textwidth]{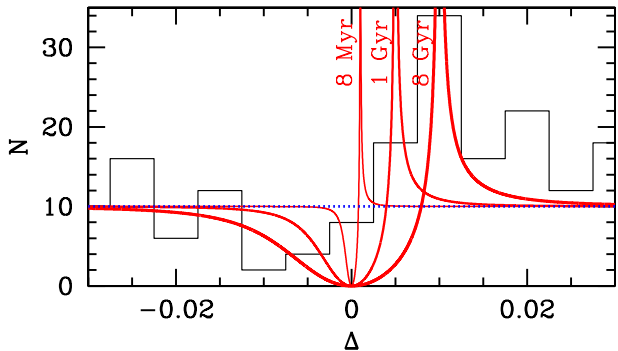}}
\caption{
 Effect of resonant repulsion on the period distribution of
planet pairs.
The black histogram shows Kepler data for planet pairs near the 2:1 and 3:2 resonances
 \citep[][]{Batalhaetal12}.
The three red curves show the effect of resonant repulsion on an initially
flat distribution of pairs at three later times.
 The parameters are chosen such that $\Delta_{\rm mig}=.005(t/{\rm Gyr})^{1/3}$
 in Equation (\ref{eq:epsilon-now}), similar to our
 fiducial values for tidal damping (Eq. (\ref{eq:eps-est})).  
 The pileup occurs at $\sim \Delta_{\rm mig}$ and the evacuated region
   extends to $\sim -\Delta_{\rm mig}$.
}
\label{fig:yfig}
\end{figure}

Two planets  that  initially have $\Delta=\Delta_0$ repel each other to
 $\Delta>\Delta_0$, and at time $t$ they  migrate to
\beqn
\Delta(t)=\left(\Delta_{\rm mig}^3+\Delta_0^3\right)^{1/3} \ ,
\label{eq:dd}
\eeqn
where
\beqn
\Delta_{\rm mig}(t) = \left( {9\over 4}\mu_1^2 \Gamma t \right)^{1/3}.
\label{eq:epsilon-now}
\eeqn
Figure \ref{fig:yfig} illustrates the effect on the distribution of period ratios.

The sign of $\Gamma$ is always positive due to eccentricity damping
alone, i.e. to the $\gamma_{ej}$ terms in Equation (\ref{eq:Gam}).
Furthermore, if tides are the source of damping, then
$\gamma_{aj}=2\gamma_{ej}$ by angular momentum conservation, leaving
$\Gamma>0$; this is also true for any form of damping that conserves
angular momentum.  Other forms of damping could in principle result in
values of $\gamma_{aj}$ that make $\Gamma$ negative.  However, the
fact that Kepler pairs are piled up outside of resonances argue that
this did not happen.

\section{Resonant Repulsion by Tides}
\label{sec:tide}

In this section we focus on the case when the dissipation is provided
by tidal damping.  The rate of eccentricity damping $\gamma_{e1}$ is
given by Equation (\ref{eq:taue}).  In addition,
$\gamma_{a1}=2\gamma_{e1}$ by angular momentum conservation, and we
may ignore tides on the outer planet ($\gamma_{e2}=\gamma_{a2}=0$)
because tidal damping rates are steep functions of orbital period.
Therefore Equation \refnew{eq:epsilon-now} becomes 
\begin{eqnarray}
\Delta_{\rm mig} 
& \approx & 0.006
\left({Q_1}\over{10}\right)^{-1/3} 
\left({k_{2}}\over{0.1}\right)^{1/3} \,
 \left({{m_1}\over{10 M_\oplus}}\right)^{1/3}\,
\left({{R_1}\over{2R_\oplus}}\right)^{5/3}\,
\nonumber \\& & 
\times \left({{M_*}\over{M_\odot}}\right)^{-8/3}\, 
\left({{P_1}\over{5 {\rm day}}}\right)^{-13/9}\,
\left({t\over{5 {\rm Gyrs}}}\right)^{1/3}\,\nonumber\\
& & \times (2\beta + 2 \beta^2)^{1/3}.
\label{eq:eps-est}
\end{eqnarray}

\begin{figure}
\centerline{\includegraphics[width=0.49\textwidth]{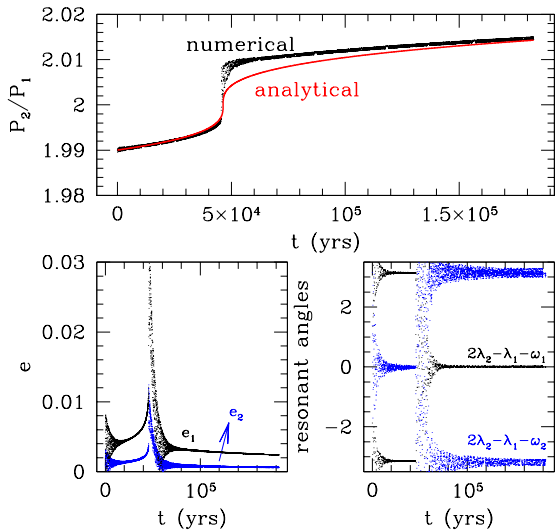}}
\caption{ An N-body simulation of resonant repulsion, where the
  dissipation is provided by tidal damping on the inner planet.  A
  pair of planets initially on the near side of the 2:1 resonance
  (period ratio $P_2/P_1=1.99$) is pushed to the far side  as a
    result of the inner planet moving inward and the outer planet
     outward.  The planets both have mass $10 M_\oplus$ and
  orbit a solar mass star, with $P_1=5$ days.  To speed up the
  simulation, we artificially enhance tides by assuming a radius of
  $12 R_E$ for the inner planet, while $Q_1 = 10$ and $k_2 = 0.1$.
  The simulation was performed with the SWIFT package
  \citep{levisonduncan}, modified to include routines for tidal
  damping and relativistic precession.
  \citep{levisonduncan}.}
\label{fig:followdelta}
\end{figure}

Figure \ref{fig:followdelta} shows an N-body simulation with tides
of two planets initially on the near side of resonance.  Resonant
repulsion pushes them to the far side, in agreement with the analytic
solution (Equations (\ref{eq:dd}) and (\ref{eq:eps-est})).  There is modest
disagreement when the pair crosses through nominal resonance when the
expansion in small $e$ becomes invalid.  The free eccentricities damp
away after a brief initial period ($\lesssim 2\times 10^4$ yr). On
crossing nominal resonance, they are regenerated, but then quickly
damp away again.  Damping locks the system into libration (of both
resonant angles), but this has little dynamical significance, as it is
merely a consequence of the eccentricities taking on their purely
forced values.

\begin{figure}[t]
\centerline{\includegraphics[width=0.48\textwidth]{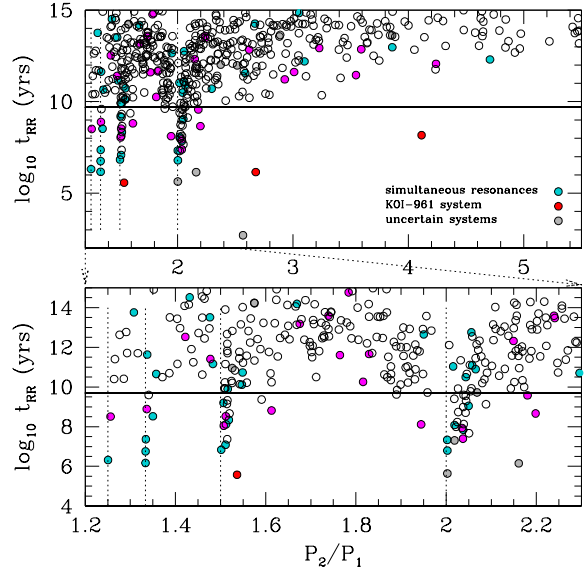}}
\caption{ 
The timescale for resonant repulsion to move the period
  ratio of Kepler planet pairs by a distance $|\Delta|$, where
  $\Delta$ is the observed fractional distance to the closest first
  order resonance.  We adopt KIC system parameters, with updated
  values for KOI-961 (red dots) from \citet{muirhead}.  For tidal
  dissipation, $Q_1 = 10$ and $k_2=0.1$.  The lower panel zooms in to
  the resonant region.  If $t_{\rm RR} \gg $ system age (the
  horizontal line is the age of the Sun), the period ratios should
  have evolved little since birth; while for $t_{\rm RR} \ll$ age, we
  do not expect the systems to linger at the observed ratios.  The
  fact that most pairs lie at or above the horizontal line is
  consistent with resonant repulsion by tides.  
  Close inspection of systems with very small $t_{\rm RR}$ reveal many
  are related to 3-body effects: the turquoise circles indicate pairs
  where one or both planets are engaged in at least two resonances
  simultaneously (defined as $|\Delta| < 3\%$). Our simple picture of
  resonant repulsion may break down in these cases.  `Uncertain
  systems' refer to those where the nominal total mass $\geq 1000
  M_\oplus$ (assuming Earth density) and we discard them from
  consideration for fear of contamination.  }
\label{fig:keplerlarge}
\end{figure}

Figure \ref{fig:keplerlarge} shows the ``resonant repulsion time''
($t_{\rm RR}$) for all reported Kepler pairs. This is the timescale
over which resonant repulsion by tides moves a pair towards or away
from the nearest first order resonance.  Mathematically, $t_{\rm
  RR}\equiv |\Delta/{\dot \Delta}|$, where $\Delta$ is the observed
fractional distance and $\dot{\Delta}$ is the rate predicted by
resonant repulsion (Equation (\ref{eq:ddelta})) assuming tidal damping is
operating with $Q_1=10$, $k_2=0.1$, and using
the observed
  planet and
stellar parameters.  On
this plot, systems that have $t_{\rm RR}$ longer than their age have
not experienced significant resonant repulsion, while all those with
shorter $t_ {\rm RR}$ should have moved to the right.

A number of inferences may be drawn.  First, most systems far from
resonances ($|\Delta| \geq 10\%$) have experienced negligible resonant
repulsion and were most likely born with the period ratio they have
today.

Second, systems within $1-10\%$ of resonance exhibit $t_{\rm RR}$ that
are as long as, or longer than, the typical age of systems (a few
Gyrs).  This is consistent with resonant repulsion by tides: systems
with shorter $t_{\rm RR}$ would have been moved to the right until
$t_{\rm RR}$ was comparable to the age of the system.  Near the 2:1
resonance, it appears that pairs as far left as $1.8$ and as far right
as $2.2$ could have been affected by the repulsion.

Last, many systems very near resonances ($|\Delta| \lesssim 1\%$)
exhibit such short $t_{\rm RR}$ that they should have migrated to much
larger $\Delta$ values.  At first sight, their presence is
troubling. However, an inspection of the Kepler catalogue reveals that
many of these are in triples or higher multiple systems, and these
planets are engaged simultaneously in two or more 2-body resonances.
The worst-off cases are  in simultaneous resonances,
reminiscent of the Laplace resonance   of Jupiter's
moons \citep{Yoder}.  Moreover, the fraction of
multiples is much higher amongst systems with $t_{\rm RR}$ falling
below the solar age line than for other random pairs.
Our simple picture of resonant repulsion fails when the planet is
subject to two or more resonances. In this case, exact resonance may
be maintained for a much longer time because the planets form a heavy
ladder with an effectively large inertia. The prevalence of
simultaneous resonances in these short $t_{\rm RR}$ systems spurs us
to hypothesize that all pairs with short $t_{\rm RR}$ in
Fig. \ref{fig:keplerlarge} are results of 3-body effects; and  that these
resonances are not primordial, but a combined effect of resonant
repulsion and 3-body effects.  

Removing the colored circles in Fig. \ref{fig:keplerlarge}, we see a
relatively clear picture that most pairs stay where they were born
with, while pairs very close to resonances experience repulsion and
are shifted by a few percent to larger period ratios.

\section{Discussion}
\label{sec:discuss}
In this work, we investigate the peculiar fact that there is an excess
of Kepler planet pairs just wide of resonance, and a deficit just
inward of resonance.  We propose that dissipation is responsible for
this asymmetry.  Two nearly resonant planets whose eccentricities are
weakly damped repel each other \citep{greenberg,lwpluto,johnpap2011}.
This is because dissipation damps away the planets' free
eccentricities, but the eccentricities that are forced by the
resonance persist despite dissipation.  Planets are typically repelled
when dissipation acts on these forced eccentricities. As such,
resonant interaction allows dissipation to continuously extract energy
from the orbits.  Resonant repulsion pushes pairs from the near side
to the far side of resonance, and naturally explains the Kepler
result.  Pairs accumulate at a fractional distance $\Delta_{\rm mig}$
wide of each resonance, with $\Delta_{\rm mig}\sim (\mu^2 t/t_{\rm
  damp})^{1/3}$, where $t_{\rm damp}$ is the typical eccentricity
damping time and $t$ the system age (Equations
(\ref{eq:dd})--(\ref{eq:epsilon-now})).

For the source of dissipation, we focused on tidal damping in the
inner planet.  The typical distance planets can repel each other is of
order a few percent or less for Kepler parameters if the tidal damping
is efficient.
The deficit of pairs immediately inward of resonance may be explained
by this repulsion. And the distances outward of resonance where planet
pairs are found are consistent with the theoretically estimated
repulsion distance. 

However, a number of inconsistencies between theory and data require
further investigation. For instance, many pairs remain very close to
resonance despite a short resonant repulsion time. 
These are often
found  in  systems with more than two planets where the planet pairs are
 engaged simultaneously in more than one resonance. We
therefore speculate that in fact all systems with short resonant
repulsion time are consequences of 3-body effects. This may be
confirmed using transit-timing variation or other tools.

If resonant repulsion is the reason behind the resonance asymmetry, 
 its signature should be observable in future studies. The
planets should currently have nearly zero free eccentricities, and as
a result both of the resonant angles should be locked at their
center-of-resonance values, with very small libration amplitude. This
can be tested with radial velocity measurements or
with transit-time variations
\citep{LXW}.
Furthermore, if
tidal damping is the dominant dissipation mechanism, we expect that
the resonance asymmetry should vanish for planets at orbital periods
greater than $10-20$ days. Long-term Kepler monitoring will decide
between tides or alternative damping mechanisms, e.g., damping by a
gaseous or  planetesimal disk.

Our study suggests that the initial period distribution of Kepler
planets was relatively flat, without major pile-ups at or near
resonances.\footnote{The chain of resonances observed in systems like
  KOI-500, KOI-730 \citep{Lissaueretal11}, KOI-2038
  \citep{Fabryckyetal12} might also not be primordial, but a combined
  result of resonant repulsion (a 2-body effect) and 3-body
  interactions. } This is in contrast to jovian mass planets and could
help constrain the origin of these low-mass planets. If disk migration
is responsible for their current location, it must somehow have
avoided pushing the planets into resonances, perhaps because the
migration rate was very fast---faster than the resonant libration
rate. Alternatively, planets may be formed {\it in-situ}
\citep{hansen} and have therefore avoided convergent migration.

\acknowledgements

We are  grateful to the Kepler team for  procuring such a
  spectacular data set.  Y.L. acknowledges support from NSF grant
AST-1109776.  Y.W. acknowledges useful conversations with J. Xie and
support from NSERC.


\end{document}